# Long-distance quantum key distribution secure against coherent attacks

B. Fröhlich, M. Lucamarini, J. F. Dynes, L. C. Comandar, W. W.-S. Tam, A. Plews, A. W. Sharpe, Z. L. Yuan, A. J. Shields

*Toshiba Research Europe Ltd, 208 Cambridge Science Park, Cambridge CB4 0GZ, UK*
*\*Corresponding author: marco.lucamarini_at_crl.toshiba.co.uk*

Quantum key distribution (QKD) permits information-theoretically secure transmission of digital encryption keys, assuming that the behaviour of the devices employed for the key exchange can be reliably modelled and predicted. Remarkably, no assumptions have to be made on the capabilities of an eavesdropper other than that she is bounded by the laws of Nature, thus making the security of QKD "unconditional". However, unconditional security is hard to achieve in practice. For example, any experimental realisation can only collect finite data samples, leading to vulnerabilities against coherent attacks, the most general class of attacks, and for some protocols the theoretical proof of robustness against these attacks is still missing. For these reasons, in the past many QKD experiments have fallen short of implementing an unconditionally secure protocol and have instead considered limited attacking capabilities by the eavesdropper. Here, we explore the security of QKD against coherent attacks in the most challenging environment: the long-distance transmission of keys. We demonstrate that the BB84 protocol can provide positive key rates for distances up to 240 km without multiplexing of conventional signals, and up to 200 km with multiplexing. Useful key rates can be achieved even for the longest distances, using practical thermo-electrically cooled single-photon detectors.

*Introduction* - Translating the exceptional properties of QKD-enabled unconditionally secure key exchange [1-3] into practice requires a rigorous approach to all aspects of the involved protocol and hardware [4-7]. Several hacking attacks on QKD systems [8-18] have shown in the past that many underlying assumptions on protocols and hardware can open the door to unwanted intrusion. However, in the presence of ideal equipment, an eavesdropper (Eve) is powerless as long as the laws of Nature are not violated. In this ideal scenario, QKD can guarantee secure communication irrespective of Eve's computational capabilities, which has been termed unconditional security.

Even so, it is sometimes convenient and insightful to study the security of a QKD protocol in presence of a limited eavesdropper. Traditionally there have been three main classes used to constrain Eve's attack: individual, collective and coherent (or general) attacks [19-24]. Only the latter does not limit the capabilities of the eavesdropper beyond what is physically possible. Any QKD system aiming to implement an unconditionally secure protocol therefore has to be proven secure against coherent attacks. Another aspect which cannot be neglected is security in a finite size scenario [25-29]. No key transmission session can be endless and the resulting statistical fluctuations have to be taken into account. For example, collective and coherent attacks against the BB84 protocol [2] are known to coincide in the asymptotic limit, whereas they can provide different secure key transmission rates in the finite-size case [30].

In the past, most QKD experiments have fallen short of the target to demonstrate security against coherent attacks in the finite size regime [31-37]. This is as much due to limitations of the employed hardware, as due to a lack of practical security proofs. For instance, only recently has the efficient BB84 protocol [38] implemented with decoy-states been proven secure against coherent attacks [39-40]. Other common protocols, such as the coherent-one-way (COW) protocol [41], are still lacking a rigorous security proof, while the differential-phase-shift (DPS) protocol has been implemented only under the assumption of individual attacks [42]. In these cases the gap between coherent and non-coherent attacks appears to be greater than in other protocols, and the maximum achievable transmission distance and key rate can be severely reduced when general attacks are considered [43, 44]. More demanding protocols based on entanglement distribution [45] or the recently developed measurement-device independent QKD [46, 47] are secure against coherent attacks, but often provide impractically low secure key rates or require a significantly more complex experimental apparatus.

In addition to security, multiplexing with conventional data communications is fundamental for the integration of QKD into existing optical networks. In the past, deployment of QKD technology has been hampered by the frequent need for dedicated "dark" fibres to segregate the very weak quantum signals from conventional traffic. The biggest limitation is the resulting broad band Raman scattering which swamps quantum signals rendering QKD inoperable. Nevertheless we have already demonstrated co-existence of QKD in short-reach environments such as quantum access networks [48] as well as high speed (> 100G) classical data traffic over distances of 100 km for metropolitan networks [49]. Here we address long haul distances and demonstrate QKD's coexistence with multiplexed classical signals over distances $\gtrsim$ 200 km.

In this article we implement two different variants of the efficient decoy-state BB84 protocol providing security against coherent attacks [39, 50] in a state-of-the-art QKD system and explore the limitations to the secure transmission distance both in dark fibre and in coexistence with conventional data transmission. We demonstrate secure key transmission over 240 km and 200 km, respectively, using highly practical thermo-electrically cooled single-photon detectors.

Figure 1 summarises results of recent long-distance QKD experiments, highlighting the significance of our result both in terms of practicality and security. Electrically cooled detectors previously have been able to support only a maximum fibre attenuation of 34 dB [33]. We extend this limit to close to 45 dB and, at the same time, demonstrate security against coherent attacks in the finite size scenario. Security against general attacks has been implemented rigorously for the efficient BB84 protocol only in one recent field trial [51] and in a proof-of-principle experiment using a plug-and-play QKD system [52], however, without a detailed analysis of the robustness and limitations in the long-distance regime. In Figure 1, the only QKD experiments performed on a distance longer than in this work are based either on the DPS protocol secure against individual attacks [34] or on the COW protocol secure against collective attacks [37]. However, the only rigorous security bounds against coherent attacks available for these protocols predict for them an attenuation range smaller than 25 dB [44]. Therefore, our results represent the longest distance currently achieved by a QKD secure against coherent attacks. It is noticeable that this is achieved in the finite-size scenario using a real QKD setup and electrically cooled detectors.

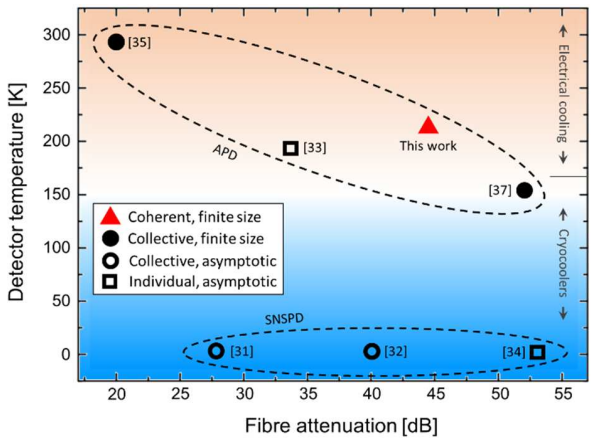

**Fig. 1**. Selection of recent long-distance QKD experiments. The graph plots the temperature of the single-photon detectors used in the experiments over the maximum attenuation that could be tolerated. Following [53], we only select demonstrations which fulfil the practical bit rate limit of > 1 bit per second (bps). Two detector types are considered, avalanche photodiodes (APDs), which can be cooled electrically, and superconducting nanowire single-photon detectors (SNSPDs), which have to be cooled cryogenically. Reference [37] uses APDs which are cooled with a cryocooler. The data points are shape coded to highlight their security level: squares – security against individual attacks; circles – collective attacks; triangles – coherent attacks. Open symbols refer to experiment considering only the asymptotic limit, whereas filled symbols take finite size effects into account.

***Coherent security*** - The protocol for our experiment is the efficient version of BB84 [38] with decoy states [54-57], featuring high key rates and unconditional security in the finite-size regime. We focus on the security proofs described in [39] and [50], which build on the min-entropy estimation from the uncertainty principle presented in [58]. Such proofs pose no assumption on the eavesdropper and assume an ideal behaviour of the equipment owned by the users. For short, we call the above two variants of the BB84 protocol "$BB84_{coh}^{(1)}$" and "$BB84_{coh}^{(2)}$", respectively, and we briefly summarise their features in the next paragraph. Also, to show the difference in the BB84 protocol's key rate when Eve performs collective or coherent attacks, we consider the BB84 protocol described in [35], called here "$BB84_{coll}$", which is secure up to the class of collective attacks. As it will become clear later, the gap between collective and coherent attacks is small for the BB84 protocol, whereas it is expected to be considerably larger for protocols like DPS and COW [44].

Both the protocols $BB84_{coh}^{(1)}$ and $BB84_{coh}^{(2)}$ use three intensity settings, $u$ ("signal"), $v$ ("decoy") and $w$ ("vacuum"), and two complementary bases, $Z$ (data basis) and $X$ (test basis) to run the decoy-state BB84 protocol. However, the two protocols present small differences that are worth mentioning as they lead to different optimisation parameters and key rates. The $BB84_{coh}^{(1)}$ protocol distils the key bits in the $Z$ basis from all the three intensity settings. The decoy-state estimation is performed using the analytical equations presented in [39] and the protocol provides the following amount of secure key bits per key session:

$$S_{coh}^{(1)} = \underline{n}_0 + \underline{n}_1[1 - h(\overline{e}_{ph})] - \lambda_{EC} - \Delta. \quad (1)$$

In Eq. (1), $\underline{n}_0$ ($\underline{n}_1$) is the lower bound to the zero-photon (single-photon) events detected in the $Z$ basis, whereas $\overline{e}_{ph}$ is the upper bound to the phase error rate of the single-photon events. The quantity $\lambda_{EC}$ is the information revealed on the public channel to correct the bit strings of the users and is directly measurable in the protocol. Finally, $\Delta = 6\log_2(21/\varepsilon_{sec}) + \log_2(2/\varepsilon_{cor})$ is a quantity related to the security and correctness of the protocol, quantified by the two parameters $\varepsilon_{sec}$ and $\varepsilon_{cor}$, respectively. For all the details about this protocol, we point the reader to Ref. [39].

The $BB84_{coh}^{(2)}$ protocol is similar to the previous protocol, but only distils the key bits from the signal intensity setting, $u$, in the basis $Z$. It adopts the linear program described in [50], numerically solved, to perform the decoy-state parameter estimation and delivers the following amount of key bits:

$$S_{coh}^{(2)} = \underline{n}_0 + \underline{n}_1 - \overline{n}_1 h(\overline{e}_{ph}) - \lambda_{EC} - \Delta'. \quad (2)$$

In Eq. (2), the quantities have analogous meaning as in Eq. (1). However, the symbol $\overline{n}_1$ represents an upper bound to the single-photon events, as opposed to the lower bound appearing in Eq. (1). This is due to a stricter interpretation of the proof method presented in [50] and would intuitively suggest a lower key rate for $BB84_{coh}^{(2)}$ than $BB84_{coh}^{(1)}$. However, this is not always the case because of the different parameter estimation routines featured by the two protocols. The quantity $\lambda_{EC}$ is the same as in Eq. (1), whereas $\Delta'$ amounts to $6\log_2(46/\varepsilon_{sec}) + \log_2(2/\varepsilon_{cor})$. In both protocols we set $\varepsilon_{sec} \lesssim 10^{-10}$ and $\varepsilon_{cor} \lesssim 10^{-15}$.

***Experimental realization*** - We explore the limits to secure key transmission for both variations of the efficient BB84 protocol secure against coherent attacks with a state-of-the-art QKD system, which is based on phase-encoding and decoy states [35, 59] (see details in [69]). We use Corning Ultra fibre with a loss coefficient of 0.18 dB/km to connect Alice (transmitter) and Bob (receiver) with a fibre distance of up to 240 km. The attenuation at this longest distance is 44.4 dB, which is higher than the expected loss of 43.2 dB due to fibre connections between several spools.

The receiver is based on thermo-electrically cooled avalanche photodiode (APD) single-photon detectors [60-62]. Electrically cooled APDs are the most practical solution to single-photon counting and they will be of particular importance for commercial applications of QKD. Other cooling methods such as Sterling refrigerators permit achieving cryogenic temperatures and therefore lower dark count rates [37], but they have a number of disadvantages. Their specified lifetime is typically no more than 5 years, they are large in size and heavy, and they are significantly more expensive [63], which makes them unlikely to be considered for telecommunication applications. At a temperature of –60°C we achieve a dark count rate of 10 counts per second (cps) at a detection efficiency of 10% with our electrically cooled APDs (Figure 2(a) and [69]). To our knowledge, this is the best noise performance of an electrically cooled single-photon detector implemented in a QKD system, and it compares well with what has been achieved with Sterling coolers [64].

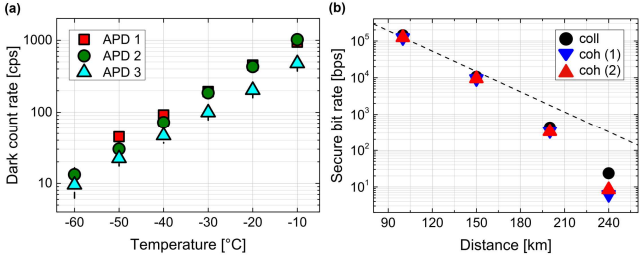

**Fig. 2**. Detector performance and secure bit rate without multiplexing of conventional signals. (a) Dark count rate as a function of APD temperature for three different devices. The dark count rate decreases by about a factor of 2 per 10°C. Error bars correspond to 1 standard deviation from 15 consecutive measurements and are in most cases smaller than the reported data points. The data points follow a colour code that suggests red (blue) shades for higher (lower) temperatures and is consistent with the colour code in Figure 3. Please see [69] for more details on single-photon detectors. (b) Secure bit rate as a function of distance for three different variants of the BB84 protocol. Black circles correspond to BB84$_{coll}$ [35], which is secure up to collective attacks in the finite-size scenario. The key rates for BB84$_{coh}^{(1)}$ [39] (downward blue triangles) and BB84$_{coh}^{(2)}$ [50] (upward red triangles) are calculated from Eqn. (1) and (2), respectively. The dashed line extrapolates the reduction of the secure key rate purely from attenuation in the fibre and provides an indication of the regime where the dark count rate plays a dominant role.

We perform a first experiment where we send only the quantum signal over the fibre channel. Figure 2(b) compares the measured secure bits $S_{coh}^{(1)}$ and $S_{coh}^{(2)}$ (Eq. (1) and (2), respectively) per second as a function of fibre transmission distance. For each distance the detector temperature and biasing conditions are adapted to achieve best performance. The decoy state intensities are set to $u = 0.5$ photons/pulse, $v = 0.11$ photons/pulse and $w = 0.0007$ photons/pulse, and the probabilities with which they are selected are $Pu = 0.5$, $Pv = 0.25$ and $Pw = 0.25$, respectively. These parameters are not optimal for all the presented protocols, but represent a compromise which permits calculating secure key rates for all protocols from exactly the same dataset. Slightly higher key rates can be achieved in some situations when optimising the parameters. Also plotted, as grey filled squares, is the secure key rate for BB84$_{coll}$ to highlight the penalty imposed by the higher security level. For the shortest distance the difference is marginal, about 10%, and it increases to more than a factor of 2 at 240 km, where we measure 23.5 bit per second (bps) and 8.4 bps, respectively. Both rates are above 1 bps, which is regarded as a practical limit for telecommunication applications [53].

The necessity to use dedicated, dark fibres to perform quantum communication would be a severe hindrance to its applicability. Coexistence of strong conventional data signals and quantum signals in the same fibre medium therefore has attracted a lot of attention in recent years [49, 65-68]. Operating QKD links over lit fibre is substantially more difficult due to excess noise generated by Raman scattering. The longest transmission distance achieved so far is 100 km [49]. Here, we explore the limits to multiplexed QKD employing a low-noise amplifier at the receiver end and narrow fibre Bragg grating based 25 GHz dense wavelength division multiplexing (DWDM) filters.

Figure 3(a) shows a schematic of the filter arrangement at the sender and receiver end (see [69] for details). We combine the quantum channel with two conventional channels with an 8-channel DWDM multiplexing module at the sender side. At the receiver, a drop filter separates the conventional signals from the quantum channel, before they are amplified with a low noise amplifier with a noise figure of < 3.3 dB. This is followed by a de-multiplexing 8-channel DWDM module separating the conventional channels. The quantum signal itself is filtered a second time using an off-the-shelf 25 GHz DWDM filter to suppress Raman noise. The additional filter increases the attenuation in Bob by approximately 3 dB. No actual data is transmitted in the experiment. One conventional channel is used to synchronise the sender and receiver with a 15.625 MHz clock signal, the other channel simulates further conventional channels with a CW laser, we refer to this channel as the simulated data channel. An increase of the launch power of the data channel can be interpreted in two ways: simulation of transmission of a higher power data channel, or transmission of an increased number of data channels [61]. An increase of 3 dB thereby corresponds to a doubling of the number of transmitted channels.

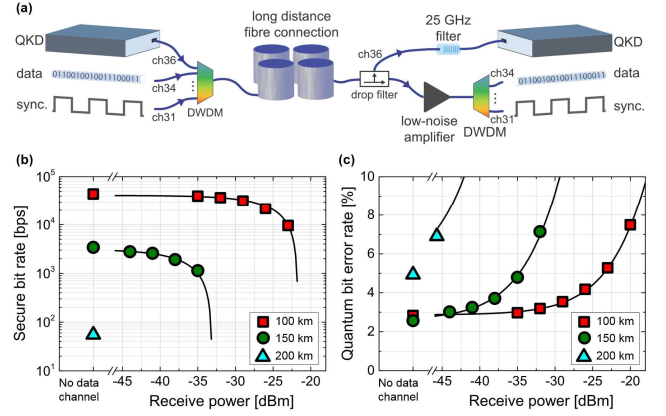

**Fig. 3**. Secure bit rate with multiplexing of conventional signals. (a) Schematic of the multiplexing setup to transmit conventional signals together with the quantum channel on the same fibre. At the transmitter side an 8-channel dense wavelength division multiplexing (DWDM) module combines the quantum signal with an optical clock and a CW laser to simulate a data channel. After transmission over the long-distance fibre, a drop filter separates the quantum signal from the conventional signals. The conventional signals are amplified before separating them with a DWDM de-multiplexer. The quantum channel is filtered with a 25 GHz DWDM filter to suppress Raman noise. Also shown are: secure bit rate (b) and quantum bit error rate (c) as a function of receive power of the simulated data channel in front of the low-noise amplifier for 100 km, 150 km, and 200 km of low-loss fibre transmission. The solid lines are the result of a simulation model (see [69]). The colour code is the same as in Figure 2, with red (blue) shades suggesting higher (lower) temperatures set in the detectors.

In Figure 3(b) and 3(c), we plot the secure key rate and quantum bit error rate (QBER), respectively, measured with the multiplexing setup as a function of receive power of the data channel in front of the amplifier. The receive power at the receiver side is an important figure of merit, as it determines if the transmitted signals can be received error-free or not. QBER and secure key rate are plotted for 100 km, 150 km, and 200 km of fibre. We plot only $S_{coh}^{(2)}$ for clarity, to avoid overlaps with $S_{coh}^{(1)}$, which is always very close to $S_{coh}^{(2)}$. The key rate secure against coherent attacks stays positive for receive powers greater than -23 dBm for a link length of 100 km. This value reduces to -35 dBm for a link length of 150 km. In a previous experiment [49] we have shown error-free operation of a 100G link together with a quantum signal for a receive power down to -35 dBm. Our results show therefore that QKD can coexist with one 100G data channel up to a distance of 150 km. However, the amplifier implemented in the setup has a specified minimum input power of -45 dBm, and it is expected that up to ten 100G channels can be multiplexed with QKD over 150 km if the receive power per channel is -45 dBm. At 200 km no additional data signal can be launched in the same fibre together with the quantum signal and the clock, even with the launch power of the clock channel set close to the minimal value for stable locking, -8.7 dBm, corresponding to a receiving power before the amplifier of -46.6 dBm. However, the transmission of the clock signal on the same fibre as the quantum signal is already an important advantage, as it improves the stability of the link.

***Conclusion*** - We have demonstrated that practical QKD systems based on thermo-electrically cooled detectors can reach transmission distances beyond 200 km, while maintaining security against the most general class of attacks allowed by quantum mechanics. This sets the current longest distance achieved by a QKD secure against coherent attacks in the finite-size scenario. Additionally, we have shown that multiplexing high-speed data signals (100G) with the quantum channel is feasible up to 150 km, while multiplexing the synchronisation signal only is feasible up to 200 km. The measured key rates compare well or exceed what has been achieved in previous demonstrations despite the higher security level. Our system is therefore ideally suited for building back-bone QKD networks, providing, for example, inter-city links. ***Acknowledgments*** - We thank Marcos Curty and Weilong Wang for useful discussions.

## REFERENCES

1. S. Wiesner, "Conjugate coding," Sigact News **15**, 78 (1983).
2. C. H. Bennett and G. Brassard, "Quantum cryptography: public key distribution and coin tossing," in Proc. IEEE Int. Conf. Comp. Systems Signal Processing 175–179 (IEEE, 1984).
3. A. K. Ekert, "Quantum cryptography based on Bell's theorem," Phys. Rev. Lett. **67**, 661 (1991).
4. N. Lütkenhaus and A. J. Shields, "Focus on quantum cryptography: theory and practice," New J. Phys. **11**, 045005 (2009).
5. V. Scarani, H. Bechmann-Pasquinucci, N. J. Cerf, M. Dušek, N. Lütkenhaus, and M. Peev, "The security of practical quantum key distribution," Rev. Mod. Phys. **81**, 1301 (2009).
6. H.-K. Lo, M. Curty, and K. Tamaki, "Secure quantum key distribution," Nat. Photon. **8**, 595 (2014).
7. E. Diamanti, H.-K. Lo, B. Qi, and Z. L. Yuan, "Practical challenges in quantum key distribution," preprint at arXiv:1606.05853 (2016).
8. G. Brassard, N. Lütkenhaus, T. Mor, and B. C. Sanders, "Limitations on practical quantum cryptography," Phys. Rev. Lett. **85**, 1330 (2000).
9. A. Vakhitov, V. Makarov, and D. Hjelme, "Large pulse attack as a method of conventional optical eavesdropping in quantum cryptography," J. Mod. Opt. **48**, 2023 (2001).
10. N. Gisin, S. Fasel, B. Kraus, H. Zbinden, and G. Ribordy, "Trojan-horse attacks on quantum-key-distribution systems," Phys. Rev. A **73**, 022320 (2006).
11. B. Qi, C.-H. F. Fung, H.-K. Lo, and X. Ma, "Time-shift attack in practical quantum cryptosystems," Quantum Inf. Comput. **9**, 73 (2005).
12. V. Makarov, A. Anisimov, and J. Skaar, "Effects of detector efficiency mismatch on security of quantum cryptosystems," Phys. Rev. A **74**, 022313 (2006).
13. C. Wiechers, L. Lydersen, C. Wittmann, D. Elser, J. Skaar, Ch. Marquardt, V. Makarov, and G. Leuchs, "After-gate attack on a quantum cryptosystem," New J. Phys. **13**, 013043 (2011).
14. I. Gerhardt, Q. Liu, A. Lamas-Linares, J. Skaar, C. Kurtsiefer, and V. Makarov, "Full-field implementation of a perfect eavesdropper on a quantum cryptography system," Nat. Commun. **2**, 349 (2011).
15. L. Lydersen, C. Wiechers, C. Wittmann, D. Elser, J. Skaar, and V. Makarov, "Hacking commercial quantum cryptography systems by tailored bright illumination," Nat. Phot. **4**, 686 (2010).
16. Y. Zhao, C.-H. F. Fung, B. Qi, C. Chen, and H.-K. Lo, "Quantum hacking: experimental demonstration of time-shift attack against practical quantum-key-distribution systems," Phys. Rev. A **78**, 042333 (2008).
17. F. Xu, B. Qi, and H. K. Lo, "Experimental demonstration of phase-remapping attack in a practical quantum key distribution system," New J. Phys. **12**, 113026 (2010).
18. J.-Z. Huang, S. Kunz-Jacques, P. Jouguet, C. Weedbrook, Z.-Q. Yin, S. Wang, W. Chen, G.-C. Guo, and Z.-F. Han, "Quantum hacking on quantum key distribution using homodyne detection," Phys. Rev. A **89**, 032304 (2014).
19. D. Mayers, "Quantum key distribution and string oblivious transfer in noisy channels," Advances in Cryptology: Proceedings of CRYPTO '96, vol. 1109, pp. 343-357 of Lecture Notes in Computer Science, N. Koblitz, Ed. (Springer-Verlag, Berlin, 1996).
20. E. Biham and T. Mor, "Security of quantum cryptography against collective attacks," Phys. Rev. Lett. **78**, 2256 (1997).
21. H.-K. Lo, and H. F. Chau, "Unconditional security of quantum key distribution over arbitrarily long distances," Science **283**, 2050 (1999).
22. P. W. Shor and J. Preskill, "Simple proof of security of the BB84 quantum key distribution protocol," Phys. Rev. Lett. **85**, 441 (2000).
23. N. Gisin, G. Ribordy, W. Tittel, and H. Zbinden, "Quantum cryptography," Rev. Mod. Phys. **74**, 145 (2002).
24. I. Devetak and A. Winter, "Distillation of secret key and entanglement from quantum states," Proc. R. Soc. A **461**, 207 (2005).
25. H. Inamori, N. Lütkenhaus, and D. Mayers, "Unconditional security of practical quantum key distribution," Eur. Phys. J. D **41**, 599 (2007).
26. M. Hayashi, "Practical evaluation of security for quantum key distribution," Phys. Rev. A **74**, 022307 (2006).
27. T. Meyer, H. Kampermann, K. Kleinmann, and D. Bruß, "Finite key analysis for symmetric attacks in quantum key distribution," Phys. Rev. A **74**, 042340 (2006).
28. M. Hayashi, "Upper bounds of eavesdropper's performances in finite-length code with the decoy method," Phys. Rev. A **76**, 012329 (2007).
29. V. Scarani and R. Renner, "Quantum cryptography with finite resources: unconditional security bound for discrete-variable protocols with one-way postprocessing," Phys. Rev. Lett. **100**, 200501 (2008).
30. M. Mertz, H. Kampermann, S. Bratzik, and D. Bruß, "Secret key rates for coherent attacks," Phys. Rev. A **87**, 012315 (2013).
31. D. Rosenberg, C. G. Peterson, J. W. Harrington, P. R. Rice, N. Dallmann, K. T. Tyagi, K. P. McCabe, S. Nam, B. Baek, R. H. Hadfield, R. J. Hughes, and J. E. Nordholt, "Practical long-distance quantum key distribution system using decoy levels," New J. Phys. **11**, 045009 (2009).
32. Y. Liu, Y. Liu, T.-Y. Chen, J. Wang, W.-Q. Cai, X. Wan, L.-K. Chen, J.-H. Wang, S.-B. Liu, H. Liang, L. Yang, C.-Z. Peng, K. Chen, Z.-B. Chen, and J.-W. Pan, "Decoy-state quantum key distribution with polarized photons over 200 km," Opt. Express **18**, 8587 (2010).
33. N. Namekata, H. Takesue, T. Honjo, Y. Tokura and S. Inoue, "High-rate quantum key distribution over 100 km using ultra-low-noise, 2-GHz sinusoidally gated InGaAs/InP avalanche photodiodes," Opt. Express **19**, 10632 (2011).
34. S. Wang, W. Chen, J.-F. Guo, Z.-Q. Yin, H.-W. Li, Z. Zhou, G.-C. Guo, and Z.-F. Han, "2 GHz clock quantum key distribution over 260 km of standard telecom fiber," Opt. Lett. **37**, 1008 (2012).
35. M. Lucamarini, K. A. Patel, J. F. Dynes, B. Fröhlich, A. W. Sharpe, A. R. Dixon, Z. L. Yuan, R. V. Penty, and A. J. Shields, "Efficient decoy-state quantum key distribution with quantified security," Opt. Express **21**, 24550 (2013).
36. P. Jouguet, S. Kunz-Jacques, A. Leverrier, P. Grangier, and E. Diamanti, "Experimental demonstration of long-distance continuous-variable quantum key distribution," Nat. Photon. **7**, 378 (2013).
37. B. Korzh, C. C. W. Lim, R. Houlmann, N. Gisin, M. J. Li, D. Nolan, B. Sanguinetti, R. Thew, and H. Zbinden, "Provably secure and practical quantum key distribution over 307 km of optical fibre," Nat. Photon. **9**, 163 (2015).

38. H.-K. Lo, H. F. Chau, and M. Ardehali, "Efficient quantum key distribution scheme and proof of its unconditional security," J. Crypt. **18**, 133 (2005).
39. C. C.-W. Lim, M. Curty, N. Walenta, F. Xu, and H. Zbinden, "Concise security bounds for practical decoy-state quantum key distribution," Phys. Rev. A **89**, 022307 (2014).
40. M. Hayashi and R. Nakayama, "Security analysis of the decoy method with the Bennett–Brassard 1984 protocol for finite key lengths," New J. Phys. **16**, 063009 (2014).
41. D. Stucki, N. Brunner, N., Gisin, V., Scarani, and H. Zbinden, "Fast and simple one-way quantum key distribution," Appl. Phys. Lett. **87**, 194108 (2005).
42. K. Inoue, "Differential phase-shift quantum key distribution systems," IEEE J. Sel. Top. Quantum Electron. **21**, 6600207 (2015).
43. J. González-Payo, F. J. Fraile-Peláez, and M. Curty, "Sequential attacks against coherent one-way quantum key distribution," Poster at the conference "QCrypt 2015", Tokyo (Japan), 28 Sep. - 2 Oct 2015.
44. T. Moroder, M. Curty, C. C. W. Lim, L. P. Thinh, H. Zbinden, and N. Gisin, "Security of distributed-phase-reference quantum key distribution", Phys. Rev. Lett. **109**, 260501 (2012).
45. T. Gehring, V. Händchen, J. Duhme, F. Furrer, T. Franz, C. Pacher, R. F. Werner, and R. Schnabel, "Implementation of continuous-variable quantum key distribution with composable and one-sided-device-independent security against coherent attacks," Nat. Commun. **6**, 8795 (2015).
46. S. L. Braunstein and S. Pirandola, "Side-channel-free quantum key distribution," Phys. Rev. Lett. **108**, 130502 (2012).
47. H.-K. Lo, M. Curty and B. Qi, "Measurement-device-independent quantum key distribution," Phys. Rev. Lett. **108**, 130503 (2012).
48. B. Fröhlich, J. F. Dynes, M. Lucamarini, A. W. Sharpe, S. W.-B. Tam, Z. L. Yuan, and A. J. Shields, "Quantum secured gigabit optical access networks," Sci. Rep. **5**, 18121 (2015).
49. J. Dynes, W. W-S. Tam, A. Plews, B. Fröhlich, A. W. Sharpe, M. Lucamarini, Z. L. Yuan, C. Radig, A. Straw, T. Edwards, A. J. Shields, "Ultra-high bandwidth quantum secured data transmission," Sci. Rep. **6**, 35149 (2016).
50. M. Lucamarini, J. F. Dynes, B. Fröhlich, Z. L. Yuan, and A. J. Shields, "Security bounds for efficient decoy-state quantum key distribution," IEEE J. Sel. Top. Quantum Electron. **21**, 6601408 (2015).
51. A. R. Dixon, J. F. Dynes, M. Lucamarini, B. Fröhlich, A. W. Sharpe, A. Plews, S. Tam, Z. L. Yuan, Y. Tanizawa, H. Sato, S. Kawamura, M. Fujiwara, M. Sasaki, and A. J. Shields, "77 day field trial of high speed quantum key distribution with implementation security," Contributed Talk at the conference "QCrypt 2016", Washington, D.C. (USA), 12 - 16 Sep. 2016.
52. F. Xu, K. Wei, S. Sajeed, S. Kaiser, S. Sun, Z. Tang, L. Qian, V. Makarov, and H.-K. Lo, "Experimental quantum key distribution with source flaws," Phys. Rev. A **92**, 032305 (2015).
53. B. Miquel and H. Takesue, "Observation of 1.5 µm band entanglement using single photon detectors based on sinusoidally gated InGaAs/InP avalanche photodiodes," New J. Phys. **11**, 045006 (2009).
54. W.-Y. Hwang, "Quantum key distribution with high loss: toward global secure communication," Phys. Rev. Lett. **91**, 057901 (2003).
55. X.-B. Wang, "Beating the photon-number-splitting attack in practical quantum cryptography," Phys. Rev. Lett. **94**, 230503 (2005).
56. H.-K. Lo, X. Ma and K. Chen, "Decoy state quantum key distribution," Phys. Rev. Lett. **94**, 230504 (2005).
57. X. Ma, B. Qi, Y. Zhao, and H.-K. Lo, "Practical decoy state for quantum key distribution," Phys. Rev. A **72**, 012326 (2005).
58. M. Tomamichel, C. C.-W. Lim, N. Gisin and R. Renner, "Tight finite-key analysis for quantum cryptography," Nat. Commun. **3**, 634 (2012).
59. L. C. Comandar, B. Fröhlich, M. Lucamarini, K. A. Patel, A. W. Sharpe, J. F. Dynes, Z. L. Yuan, R. V. Penty, and A. J. Shields, "Room temperature single-photon detectors for high bit rate quantum key distribution," Appl. Phys. Lett. **104**, 021101 (2014).
60. M. A. Itzler, X. Jiang, M. Entwistle, K. Slomkowski, A. Tosi, F. Acerbi, F. Zappa, and S. Cova, "Advances in InGaAsP-based avalanche diode single photon detectors," J. Mod. Opt. **58**, 174 (2011).
61. J. Zhang, M. A. Itzler, H. Zbinden and J.-W. Pan, "Advances in InGaAs/InP single-photon detector systems for quantum communication," Light Sci. Appl. **4**, e286 (2015).
62. L. C. Comandar, B. Fröhlich, J. F. Dynes, A. W. Sharpe, M. Lucamarini, Z. L. Yuan, R. V. Penty, and A. J. Shields, "Gigahertz-gated InGaAs/InP single-photon detector with detection efficiency exceeding 55% at 1550 nm," J. Appl. Phys. **117**, 083109 (2015).
63. R. Radebaugh, "Cryocoolers: the state of the art and recent developments," J. Phys. Cond. Mat. **21**, 164219 (2009).
64. B. Korzh, N. Walenta, T. Lunghi, N. Gisin and H. Zbinden, "Free-running InGaAs single photon detector with 1 dark count per second at 10% efficiency," Appl. Phys. Lett. **104**, 081108 (2014).
65. P. D. Townsend, "Simultaneous quantum cryptographic key distribution and conventional data transmission over installed fibre using wavelength-division multiplexing," Electron. Lett. **33**, 188 (1997).
66. T. E. Chapuran, P. Toliver, N. A. Peters, J. Jackel, M. S. Goodman, R. J. Runser, S. R. McNown, R. Dallmann, R. J. Hughes, and K. P. McCabe, "Optical networking for quantum key distribution and quantum communications," New J. Phys. **11**, 105001 (2009).
67. P. Eraerds, N. Walenta, M. Legré, N. Gisin, and H. Zbinden, "Quantum key distribution and 1 Gbps data encryption over a single fibre," New J. Phys. **12**, 063027 (2010).
68. K. A. Patel, J. F. Dynes, I. Choi, A. W. Sharpe, A. R. Dixon, Z. L. Yuan, R. V. Penty, and A. J. Shields, "Coexistence of high-bit-rate quantum key distribution and data on optical fiber," Phys. Rev. X **2**, 041010 (2012).
69. Supplementary Materials.